# Player Chemistry: Striving for
# a Perfectly Balanced Soccer Team

Lotte Bransen[1] – Jan Van Haaren[1,2]
[1]SciSports, The Netherlands – [2]KU Leuven, Belgium
{l.bransen, j.vanhaaren}@scisports.com

## 1. Introduction

In the dying hours of the 2010/2011 winter transfer window, Liverpool signed Andy Carroll from Newcastle United. Although *The Reds* had to break their transfer record, Carroll never met the expectations. The forward, who had scored 11 goals in 19 matches for *The Magpies* in the first half of the season, managed just 6 goals from 44 appearances for Liverpool. The then-22-year-old Carroll is a tall target man thriving from crosses, but Liverpool did not play that way and also did not have the appropriate players in their squad to accommodate his preferred playing style. In contrast, the more mobile Luis Suárez, who was signed on the same day and gelled much better with the players surrounding him on the pitch, scored 69 goals in 110 matches for Liverpool.

Although this example illustrates the importance of chemistry among players, recruitment departments at soccer clubs still largely ignore this aspect in their evaluation of potential signings. Typically, scouts solely focus on the individual qualities of potential signings and do not account for the team balance and team chemistry in their judgments on the players. On one hand, assessing the potential chemistry with the players who are already on their team is hard by watching a limited number of matches either on tape or live in the stadium. On the other hand, existing soccer analytics tools that can process large volumes of match data mostly focus on evaluating the individual performances of players. However, the topic of assessing the impact of interactions between players on team performance is receiving increasing attention in soccer (e.g., [1], [2]) and basketball (e.g., [3], [4], [5]), which are both highly dynamic cooperative sports.

This paper takes a first step towards objectively providing insight into the question: How well does a team of soccer players gel together? We address this question in both an observational and a predictive setting. In the former setting, we *observe* the chemistry between players who have actually played together. This setting is relevant for a manager who needs to decide on the best possible line-up for an upcoming match. In the latter setting, we *predict* the chemistry between players who have never played together before. This setting is particularly relevant for a scout who needs to assess the fit of a potential signing with the players who are already on the team. In order to do so, we adopt the assumption that two players who have high mutual chemistry perform better than players who have low mutual chemistry, all other things being equal.

In this paper, we present the following four contributions:

1. We introduce two chemistry metrics that measure the offensive and defensive chemistry for a pair of players. The offensive chemistry metric measures the pair's joint performance in terms of scoring goals, whereas the defensive chemistry metric measures their joint performance in preventing their opponents from scoring goals.
2. We introduce two machine-learned models that predict the offensive and defensive chemistry for a pair of players that has never played together in a match.
3. We propose a Team Builder that automatically assembles a team of eleven players from a set of players that maximizes the mutual chemistries between the players on the team.
4. We present three observations that show the utility of our chemistry metrics as well as three concrete use cases that demonstrate the practical applicability of our chemistry metrics and Team Builder.

We compute our chemistry metrics for 361 seasons in 106 different competitions and show that the partnership between Mohamed Salah and Roberto Firmino in Liverpool's 2017/2018 UEFA Champions League campaign exhibited the highest mutual chemistry between two players. We also show that Mesut Özil's chemistry has rapidly started declining following Alexis Sánchez' departure to Manchester United in 2018. Furthermore, we identify Bayern Munich as the best-suited next destination for Moroccan international Hakim Ziyech in terms of chemistry if the attacker would decide to leave Dutch champions Ajax. Using our Team Builder, we also assemble the maximum-chemistry team for a Bayern Munich side that features Hakim Ziyech as a right winger.

The remainder of this paper is organized as follows. Section 2 describes our dataset. Section 3 presents our metrics for *measuring* the mutual chemistry for a pair of players that has played together in matches, and Section 4 presents our approach for *predicting* the mutual chemistry for a pair of players that has never played together in matches. Section 5 introduces our Team Builder that automatically assembles an optimal-chemistry team from a given set of players. Section 6 presents three observations from analyzing our dataset, and Section 7 presents three concrete applications of our chemistry metrics and Team Builder. Section 8 concludes the paper and discusses directions for future work.

## 2. Dataset

Our dataset comprises Wyscout match event data for a selection of 106 domestic and international competitions since the start of the 2015/2016 season, which describe the on-the-ball actions that the players performed in each match. While soccer data comes in many different flavors, we use match event data for developing our mutual chemistry metrics. With match event data being available for most professional football competitions around the world, this type of data strikes a good balance between data availability and data granularity. Hence, match event data is ideally suited for developing performance and style metrics that are aimed at player recruitment.

We convert our match event data into the SPADL[1] representation to facilitate our analysis [6]. Hence, for each on-the-ball action, our dataset contains a match identifier, a team identifier, a player identifier, the time in the match, the start and end location, the result (e.g., successful or unsuccessful), and the body part that the player used to perform the action. In addition, our dataset contains personal information about the players and coaches who participated in the matches in our dataset. For each person, our dataset contains their age, height, weight, nationality, mother tongue, region (e.g., Americas), subregion (e.g., South America), preferred foot, and position.

---

[1] We use the *socceraction* Python package: https://github.com/ML-KULeuven/socceraction/.

Furthermore, we enrich our dataset in two ways. First, we obtain ability and playing style indicators for each player from SciSports. We use SciSkill and Potential to reflect the current and potential ability of each player [7]. We use Player Roles to describe the playing style of each player [8]. We use Physical Performance Indicators to capture the duel strength, speed and work rate of each player.[2] Second, for each pair of players, we derive whether they were born in the same region, speak the same language and played in the same competition ahead of each season. We also compute the number of matches that they had played together ahead of each season for both club and country. Table 1 presents the characteristics of our extensive dataset.

**Table 1.** Our extensive dataset covers 361 seasons, 106,496 matches, 2,154 teams and 38,447 players from 106 domestic and international competitions across the world.

| Property | Count |
|---|---:|
| Competitions | 106 |
| Seasons | 361 |
| Matches | 106,496 |
| Teams | 2,154 |
| Players | 38,447 |
| Unique player - season combinations | 97,491 |
| Unique nationalities of players | 190 |
| Unique mother tongues of players | 68 |

## 3. Measuring the mutual chemistry for a pair of players

We exploit the observation that the mutual chemistry between players is reflected in their performances. Since we cannot observe the mutual chemistry between two players directly, we adopt the assumption that two players who have high mutual chemistry perform better than two players who have low mutual chemistry, all other things being equal. Hence, we need a performance metric that measures the joint performance for a pair of players. Since no such metric exists to the best of our knowledge, we introduce two novel performance metrics that measure the joint performance for a pair of players. We build on top of the VAEP framework for valuing individual on-the-ball actions of players [6], which was inspired by the EPV framework for basketball [9]. The VAEP framework rewards each on-the-ball player action based on its impact on the player's team's chances of scoring and conceding a goal, where the reward can be positive or negative. We present one metric that captures the joint offensive impact and one metric that captures the joint defensive impact. We now discuss each of both metrics in turn.

---

[2] SciSports developed four metrics based on match event data that assess the air duel strength, ground duel strength, speed and work rate of players on a five-point scale. This research has not been published yet.

## 3.1. Joint Offensive Impact (JOI)

We introduce the JOI metric to capture the ***Joint Offensive Impact*** for a pair of players. Intuitively, JOI quantifies the impact of the actions in which both players are involved on increasing the likelihood of scoring a goal. For example, when a pass moves the ball into a more dangerous pitch location, both the sender and the receiver will receive credit for the pass.

To obtain the JOI for a pair of players in a season, we need to address two tasks. First, we extend the notion of the VAEP rating from capturing the impact of a single action to capturing the impact of an interaction, which are two consecutive actions where each of both players performed one action each. For example, when a player passes the ball to a teammate and this teammate subsequently takes on his direct opponent, then the sequence of these two actions is considered an interaction. The sum of the VAEP ratings for the pass and the take-on reflect the impact of this interaction. Second, we compute the JOI for a pair of players in a match by aggregating the VAEP ratings for their interactions in that match.

### 3.1.1. Computing the offensive impact for interactions

We view a match as a sequence of actions $\{a_1^p, \ldots, a_n^p\}$, where $n$ is the number of actions in the match and $p \in P$ refers to the player who performed the action with $P$ being the set of all players in the match. Each action $a_i^p$ has one of the following five types: pass, cross, dribble, take-on, or shot. Formally, we define the $j$th interaction between two players $p$ and $q$ in a match $m$ as a subsequence of two consecutive actions as follows: $I_j^m(p, q) = (a_i^p, a_{i+1}^q)$, where $a_i^p$ represents the $i$th action in a match $m$ performed by a player $p$.

Furthermore, we extend the notion of the VAEP rating from actions to interactions between players. We obtain the VAEP rating for an interaction by summing the VAEP ratings for the two constituting actions. Formally, we define this extension as $VAEP(I_j^m(p, q)) = VAEP(a_i^p) + VAEP(a_{i+1}^q)$.

### 3.1.2. Computing the joint offensive impact in a season

We compute the joint offensive impact for a pair of players in a match by summing the VAEP ratings for their interactions in that match, regardless of which of both players initiated the interaction. Formally, we define $JOI_m(p, q)$ for a pair of players $(p, q)$ in a match $m$ as follows:

$$JOI_m(p, q) = \sum_k VAEP(I_k^m(p, q)) + \sum_l VAEP(I_l^m(q, p)),$$

where $k$ is the number of interactions between players $p$ and $q$ where player $p$ performs the first action and player $q$ performs the second action, and $l$ is the number of interactions between these players where player $q$ performs the first action and player $p$ performs the second action.

Furthermore, to allow a fair comparison between pairs of players who spent different amounts of time on the pitch together within a season, we introduce $JOI90(p, q)$ that represents the Joint Offensive Impact per 90 minutes of play for a pair of players $(p, q)$ in that season. Formally, we define $JOI90(p, q)$ as follows:

$$JOI90(p, q) = \sum_m JOI_m(p, q) * \frac{90}{\sum_m MINS_m(p, q)},$$

where we sum over all matches $m$ in a season and $MINS_m(p, q)$ represents the number of minutes that players $p$ and $q$ spent together on the pitch in match $m$.

## 3.2. Joint Defensive Impact (JDI)

We introduce the JDI metric to capture the ***Joint Defensive Impact*** for a pair of players. Intuitively, JDI quantifies the impact of the actions in which both players are involved on decreasing the likelihood of conceding a goal. Unfortunately, the match event data only describes the actions that actually happened in the match but not the actions that players prevented from happening, for instance, by smart runs or clever positioning. To overcome this restriction of the match event data, the rationale behind our JDI metric is that when an opponent underperforms their expected offensive impact, then the pairs of players that are responsible for defending this opponent likely have done well. We assume that a pair of players is good at preventing their opponents from making an offensive impact when their opponents structurally underperform their expected offensive impacts. We also assume that multiple pairs of players can be responsible for defending the same opponent.

We build upon the notion of the VAEP rating to capture how well a pair of players prevents their direct opponents from having an offensive impact. That is, we compare the offensive impact that a particular player is expected to have in a match with the offensive impact that this player actually had in the match. We distribute the credit for the difference between the expected and actual offensive impact for each player across the pairs of players that were responsible for preventing the player from being dangerous. For example, a central defender and right back will be responsible for preventing the opponent's left winger to be dangerous in many situations throughout a match.

To obtain the JDI for a pair of players, we need to address three tasks. First, we compute the offensive impact above or below expectation for each opponent. Second, we compute the responsibility share for the combination of player pair and opponent. For example, a player pair consisting of a central defender and the right back will have a higher responsibility share than a player pair consisting of a central defender and left back to prevent the opponent's left winger from having an offensive impact. Third, we compute the JDI for a pair of players in a match by combining the offensive impacts above or below expectation and the responsibility shares for that match.

### 3.2.1. Computing the offensive impact above or below expectation for each opponent

We obtain the offensive impact above or below expectation for each opponent in a match by computing the difference between the expected offensive impact in the match based on earlier matches in the season and the actual offensive impact. An opponent who achieved a *higher* actual offensive impact than expected offensive impact has performed *above* expectation. In contrast, an opponent who achieved a *lower* actual offensive impact than expected offensive impact has performed *below* expectation.

We determine the *actual* offensive impact for a player in a match by summing the VAEP ratings for the player's passes, crosses, dribbles, take-ons, and shots in that match. Formally, we compute the *actual* offensive impact for a player $p$ in a match $m$ as follows:

$$OI_m(p) = \sum_k VAEP(a_k^p),$$

where each action $a_k^p$ is a pass, cross, dribble, take-on, or shot performed by player $p$ in match $m$.

We determine the *expected* offensive impact for a player in a match by computing the player's average offensive impact in the matches in the same season of the same competition that were played earlier. For example, to determine the *expected* offensive impact of Manchester City's Kevin De Bruyne in the match against Liverpool on matchday 12 of the 2019/2020 Premier League season, we consider De Bruyne's performances in the first 11 Premier League matches. Formally, we compute the *expected* offensive impact for a player $p$ in a match $m$ as follows:

$$E\left[OI_m(p)\right] = \sum_1^{m-1} OI_m(p) * \frac{90}{\sum_1^{m-1} MINS_m(p)},$$

where $MINS_m(p)$ represents the number of minutes player $p$ played in match $m$.

However, we apply a Bayesian approach to obtain more robust estimations of the *expected* offensive impact for players who have played fewer than 700 minutes, for instance at the start of a season or due to an injury. We compute a weighted average between a position-specific prior and the earlier computed *expected* offensive impact, where the weights are linearly proportional to the share of the required 700 minutes the player has played. For example, for a player who has played exactly 350 minutes, the prior and computed *expected* offensive impact contribute equally to the final expectation. We determine each position-specific prior by computing the average offensive impact per 90 minutes for players on that position.

### 3.2.2. Computing the responsibility share for each pair of players and opponent

Each pair of players is responsible for preventing their opponents from having an offensive impact. However, the degree to which a pair of players is responsible for guarding a given opponent depends on the positions of the three involved players. Since we do not know the exact spatial locations of the players at the time of each action, we use the default positions of the players to determine a *responsibility share* for each combination for a pair of players and an opponent. For example, the pair consisting of the right central defender and the right back will have a large responsibility share in preventing the opponent's left winger from having an offensive impact.

We compute the responsibility share for each combination for a pair of players and an opponent in two steps. First, we compute the individual responsibility share for each individual player and a given opponent by computing the Euclidean distance between their positions. We overlay the pitch with the 5-by-5 grid in Table 2 from the perspective of each team and compute the distance between the positions in this grid. The responsibility share is inversely proportional to the Euclidean distance in the grid. The lower the distance is, the higher the responsibility share is. Second, to obtain the responsibility share for a pair, we compute the average of the individual responsibility shares of the two players that constitute the pair. Formally, we define the responsibility share for a pair of players $(p, q)$ for an opponent $o$ in a match $m$ as follows:

$$RESP_m(p, q, o) = \frac{RESP_m(p, o) + RESP_m(q, o)}{2},$$

where $RESP_m(p, o)$ represents the individual responsibility share of player $p$ for defending opponent $o$ in match $m$ based on the Euclidean distance between their positions in the grid.

**Table 2.** We obtain the responsibility share for each combination of player pair and opponent by computing the Euclidean distance between the positions[3] of the players using this 5-by-5 grid.

| Left Wing Forward | | Striker | | Right Wing Forward |
|---|---|---|---|---|
| | Left Attacking Midfielder | Second Striker & Attacking Midfielder | Right Attacking Midfielder | |
| Left Winger | Left Center Midfielder | | Right Center Midfielder | Right Winger |
| Left Wingback | Left Defensive Midfielder | Defensive Midfielder | Right Defensive Midfielder | Right Wingback |
| Left Back | Left Center Back | Center Back | Right Center Back | Right Back |

### 3.2.3. Computing the joint defensive impact in a season

We compute $JDI_m(p, q)$ for a pair of players $(p, q)$ in a match $m$ by summing the differences between the expected offensive impact and actual offensive impact for each opponent, weighted by the responsibility shares and accounted for the number of minutes spent together on the pitch. Formally, we define $JDI_m(p, q)$ as follows:

$$JDI_m(p, q) = \sum_o \left[ (E\left[OI_m(o)\right] - OI_m(o)) * RESP_m(p, q, o) * \frac{MINS_m(p,q,o)}{90} \right],$$

where $MINS_m(p, q, o)$ represents the number of minutes that players $p$ and $q$, and opponent $o$ spent together on the pitch in match $m$.

Furthermore, to allow a fair comparison between pairs of players who spent different amounts of time on the pitch together in a season, we introduce $JDI90(p, q)$ that represents the Joint Defensive Impact per 90 minutes of play for a pair of players $(p, q)$ in that season. Formally, we define $JDI90(p, q)$ as follows:

---

[3] We base our mapping on the player positions that Wyscout uses in their match event data: https://apidocs.wyscout.com/matches-wyid-events#6-details-of-formations-object.

$$JDI90\,(p,\,q) \;=\; \sum_m JDI_m(p,\,q)\;*\;\frac{90}{\sum_m MINS_m\,(p,\,q)}\,,$$

where we sum over all matches $m$ in a season and $MINS_m(p,\,q)$ represents the number of minutes that players $p$ and $q$ spent together on the pitch in match $m$.

# 4. Predicting the mutual chemistry for a pair of players

When two players have spent a sufficient number of minutes on the pitch together, the JOI and JDI metrics provide useful insights in their mutual chemistry. A manager could, for instance, use these insights to decide on the strongest possible line-up for a match. However, the JOI and JDI metrics cannot be used directly for players who have not played together often enough yet. To overcome this limitation, we introduce machine-learned models that predict the joint offensive impact and joint defensive impact for any given pair of players, regardless of whether these players have ever played together. A scout could, for instance, use these predictions to assess whether a potential signing would be a good fit in terms of chemistry with the players who are already on the team.

## 4.1. Representing players as feature vectors

We represent each player in our dataset as a feature vector that captures the characteristics of the player that likely influence the player's chemistry with another player. These characteristics include the player's age, position line (e.g., goalkeeper, defender, midfielder, or striker), height, weight, nationality, region (e.g., Europe), subregion (e.g., Western Europe), mother tongue, Physical Performance Indicators(i.e., ground duel strength, air duel strength, speed, and work rate) and scores for each of the 22 Player Roles.

Similarly, we represent each player pair in our dataset as a feature vector that combines the characteristics of the players who constitute the pair. Furthermore, the feature vector captures whether the players have the same nationality, have the same mother tongue, and originate from the same region and subregion. In addition, the feature vector also captures the number of matches the players have played together before ahead of the season.

## 4.2. Predicting the joint impact for a pair of players

We train two machine-learned regression models to predict the joint impact for a pair of players in a given season. One model predicts the joint offensive impact, whereas the other model predicts the joint defensive impact. To do so, we split our dataset into a training set, validation set and test set as is common in machine learning. Our training set covers the 2015/2016 through 2016/2017 seasons, our validation set covers the 2017/2018 season and our test set covers the 2018/2019 and 2019/2020 seasons until Tuesday 10 December 2019, for which we report observations and present use cases later.[4] We construct one example for each pair of players that played at least 700 minutes together. As a result, our training set contains 355,671 pairs, our validation set contains 185,927 pairs, and our test set contains 234,408 pairs. We use the feature representation from Section 4.1 to represent the player pairs. We use the JOI90 metric as the label to train the model that predicts the joint offensive impact, whereas we use the JDI90 metric as the label to train the model that predicts the joint defensive impact.

---

[4] We include the data for competitions that are on a calendar-year cycle (e.g., MLS) to the same set as the data for competitions that are on a season cycle (e.g., English Premier League) based on the year in which the season ends. For example, the data for the 2017/2018 season and 2018 calendar year are in the same set.

We train the models using the CatBoost[5] gradient boosting toolkit [10], which handles categorical features (e.g., region or subregion) in a natural way. We train the models on the training set, tune the hyperparameters on the validation set, and select the appropriate features by optimizing the Root Mean Square Error (RMSE) on the validation set. The model that predicts the joint offensive impact consists of 500 trees with a maximum depth of 7, while the model that predicts the joint defensive impact consists of 1000 trees with a maximum depth of 5. We compare the performance of both models on the test set to a baseline that predicts the average JOI90 or JDI90 in the training set for each example. Our models both outperform this baseline as can be seen in Table 3.

**Table 3.** Root Mean Square Error (RMSE) on the test set for both our models and the baselines. In both settings, our models outperform the baseline models.

|  | RMSE for the baseline | RMSE for our model |
|---|---|---|
| Joint offensive impact | 0.05448 | 0.04464 |
| Joint defensive impact | 0.89075 | 0.88906 |

### 4.3. Analyzing the machine-learned models

To gain a better understanding of which characteristics contribute most to the chemistry between a pair of players, we investigate the importance of each feature in both machine-learned models with respect to making accurate predictions. In both cases, the scores for a number of Player Roles have a considerable impact on the predictions. The scores for the Mobile Striker, Deep-Lying Playmaker and Ball-Playing Defender roles contribute to the predictions of the JOI90 metric, while the scores for the Holding Midfielder and Ball-Winning Defender roles do so for the JDI90 metric.

Furthermore, the number of matches that two players had played together ahead of the season has a considerable impact on the predictions of both the JOI90 and JDI90 metric for a pair of players. This effect is most notable for players who have not or have barely played together. However, their mutual chemistry can quickly increase after playing a few matches together. Beyond the mark of 50 matches together (i.e., slightly more than a season in most competitions), this effect is much weaker.

Somewhat surprisingly, the cultural features (e.g., whether the players have the same nationality, whether they originate from the same region, or whether they have the same mother tongue), exhibit only limited predictive power in both models. Supposedly, it is more important that two players communicate well with their feet and that they have complementary playing styles.

## 5. Automatically assembling a maximum-chemistry team

The ability to estimate the offensive and defensive chemistry between any pair of players enables several different applications. One potentially exciting use case for a soccer club's manager or recruitment department is to automatically assemble a maximum-chemistry team from a set of players. In addition to analyzing traditional performance metrics and style indicators, a recruitment department can also investigate the likely chemistry between the players who are already on the team and each of the transfer targets on their shortlist. To facilitate this use case, we introduce our Team Builder which automatically assembles a maximum-chemistry team from a set of players.

Inspired by the approach taken in [1], our Team Builder addresses this combinatorial optimization task by solving a mixed-integer programming problem where the goal is to select precisely eleven players who maximize the sum of their mutual offensive and defensive chemistry.[6] We allow the user to trade-off offensive chemistry for defensive chemistry by providing a weighting parameter in the optimization. Formally, we solve the following mixed-integer programming problem:

$$max \sum_{p \in P} \sum_{q \in P} (\alpha * E[JOI90\,(p,q)] \; + \; (1-\alpha) * E[JDI90\,(p,q)]) \; * x_p * x_q$$

$$subject\ to \quad \sum_{p \in P} x_p = 11\,,$$

$$\sum_{p \in P} (GKP_p * x_p) = 1\,,$$

$$3 \leq \sum_{p \in P} (DEF_p * x_p) \leq 5\,,$$

$$3 \leq \sum_{p \in P} (MID_p * x_p) \leq 5\,,$$

$$1 \leq \sum_{p \in P} (FWD_p * x_p) \leq 3\,,$$

where each $x_p$ is a decision variable that indicates whether the player $p$ is selected for the team or not, and the $GKP_p$, $DEF_p$, $MID_p$, and $FWD_p$ indicators indicate whether the player $p$ is a goalkeeper, defender, midfielder, or forward, respectively. The first constraint ensures that precisely 11 players are selected, while the other constraints ensure that a logical formation is constructed.

## 6. Observations

We now present a number of observations that result from computing the Joint Offensive Impact (JOI) and Joint Defensive Impact (JDI) metrics for all seasons in our dataset. First, we present three lists of highest-chemistry player pairs. Second, we analyze the chemistry of the Liverpool team that won the 2018/2019 Champions League. Third, we analyze the mutual chemistry between Mesut Özil and his Arsenal teammates since the start of the 2015/2016 season.

---

[6] We use the *PuLP* Python package: https://github.com/coin-or/pulp.

## 6.1. Highest-chemistry player pairs

Table 4 shows the top-ten-ranked player pairs in terms of Joint Offensive Impact per 90 minutes (JOI90) across all seasons in our dataset. Mohamed Salah and Roberto Firmino top the ranking with their partnership at Liverpool in the 2017/2018 Champions League. Luis Suárez and Lionel Messi in the 2015/2016 Primera División season rank second. Female players Yuki Nagasato and Sam Kerr complete the top three with their joint performances in the 2019 NWSL season.

**Table 4.** Player pairs with the highest Joint Offensive Impact per 90 minutes across all seasons in our dataset with at least 900 minutes played together.

| # | Player pair | Team | Season | Competition | JOI90 |
|---|---|---|---|---|---|
| 1 | Mohamed Salah Roberto Firmino | Liverpool | 2017/2018 | UEFA Champions League | 0.7077 |
| 2 | Luis Suárez Lionel Messi | Barcelona | 2015/2016 | Primera División (Spain) | 0.6497 |
| 3 | Yuki Nagasato Sam Kerr | Chicago Red Stars | 2019 | NWSL (USA) | 0.6096 |
| 4 | Artem Milevskiy Pavel Nekhaychik | Dinamo Brest | 2017 | Premier League (Belarus) | 0.6050 |
| 5 | Oleksandr Mishurenko Oleksandr Akymenko | Inhulets | 2018/2019 | Persha Liga (Ukraine, 2nd tier) | 0.6017 |
| 6 | Zlatan Ibrahimović Ángel di Maria | PSG | 2015/2016 | Ligue 1 (France) | 0.5972 |
| 7 | Dario Kovačić Sirlord Conteh | St. Pauli | 2018/2019 | Regionalliga (Germany, 4th tier) | 0.5873 |
| 8 | Gabriel Barbosa Giorgian de Arrascaeta | Flamengo | 2019 | Serie A (Brazil) | 0.5763 |
| 9 | Leroy Sané Raheem Sterling | Manchester City | 2018/2019 | Premier League (England) | 0.5701 |
| 10 | Dusan Tadić Quincy Promes | Ajax | 2019/2020 | Eredivisie (Netherlands) | 0.5700 |

Table 5 shows the top-five-ranked player pairs in terms of JOI90 in the 2018/2019 and 2019 seasons for a selection of leagues. In the German Bundesliga, right winger Karim Bellarabi gels well with striker Kevin Volland at Bayer Leverkusen. In the Major League Soccer (MLS), Jozy Altidore and Alejandro Pozuelo top the list, which also did not go unnoticed by the media.[7] In the National Women's Soccer League (NWSL), Chicago Red Stars' Yuki Nagasato and Sam Kerr, who led their team to the NWSL final, gel best.

**Table 5.** Player pairs with the highest Joint Offensive Impact per 90 minutes in the 2018/2019 season with at least 900 minutes played together or 450 minutes for the Champions League.

| # | German Bundesliga 2018/2019 | Champions League 2018/2019 | American NWSL 2019 | American MLS 2019 | Mexican Liga MX 2018/2019 |
|---|---|---|---|---|---|
| 1 | Kevin Volland Karim Bellarabi | Kylian Mbappé Neymar | Yuki Nagasato Sam Kerr | Jozy Altidore Alejandro Pozuelo | Lucas Zelarayán André-Pierre Gignac |
| 2 | Serge Gnabry Thomas Müller | Luis Suárez Lionel Messi | Crystal Dunn Lynn Williams | Johnny Russell Krisztián Németh | Alexis Vega Isaac Brizuela |
| 3 | Jadon Sancho Marco Reus | Lorenzo Insigne José Callejón | Debinha Lynn Williams | Carles Gil Gustavo Bou | Pablo López Franco Jara |
| 4 | Marco Reus Mario Götze | Hakim Ziyech Dusan Tadić | Cristine Sinclair Tobin Heath | Cristian Pavón Zlatan Ibrahimović | Miler Bolaños Fabián Castillo |
| 5 | Lucas Alario Kevin Volland | Dusan Tadić Klaas-Jan Huntelaar | Morgan Brian Sam Kerr | Brian White Alejandro Romero Gamarra | Victor Guzmán Edwin Cardona |

Table 6 shows the top-five-ranked player pairs in terms of Joint Defensive Impact per 90 minutes. In the Champions League, Tottenham Hotspurs' right back Kieran Trippier gels well with midfielders Son Heung-Min and Christian Eriksen. In the Mexican Liga MX, Leon's experienced center backs Ramiro González and William Tesillo top the list.

---

[7] https://torontosun.com/sports/soccer/mls/altidore-pozuelo-form-great-duo-for-tfc

**Table 6.** Player pairs with the highest Joint Defensive Impact per 90 minutes in the 2018/2019 season with at least 900 minutes played together or 450 minutes for the Champions League.

| # | German Bundesliga 2018/2019 | Champions League 2018/2019 | American NWSL 2019 | American MLS 2019 | Mexican Liga MX 2018/2019 |
|---|---|---|---|---|---|
| 1 | Nordi Mukiele Marcel Halstenberg | Son Heung-Min Kieran Trippier | Lauren Barnes Theresa Nielsen | Auro Omar González | William Tesillo Ramiro González |
| 2 | Christian Günter Lukas Kübler | Kieran Trippier Christian Eriksen | Theresa Nielsen Beverly Goebel-Yanez | Maynor Figueroa Adolph DeLaGarza | Celso Ortiz José María Basanta |
| 3 | Miloš Veljković Niklas Moisander | Fabinho Trent Alexander-Arnold | Megan Oyster Lauren Barnes | Adam Lundqvist Adolph DeLaGarza | Carlos Salcedo Rafael Carioca |
| 4 | Miloš Veljković Nuri Şahin | Giorgio Chiellini Leonardo Bonucci | Megan Oyster Theresa Nielsen | Chris Mavinga Omar González | Jorge Torres Rafael Carioca |
| 5 | Jeffrey Gouweleeuw Martin Hinteregger | Dani Olmo Marin Leovac | Sarah Gorden Sam Kerr | Justin Meram Leandro González Pírez | Javier Salas Adrián Aldrete |

Due to space restrictions, we only present the top-five-ranked player pairs in terms of both Joint Offensive Impact per 90 minutes and Joint Defensive Impact per 90 minutes for five competitions. However, the full results for all seasons in our dataset are available as an online appendix.[8]

## 6.2. Mutual chemistry for the 2018/2019 Liverpool side

Liverpool fans will not quickly forget the 2018/2019 season. The side of manager Jürgen Klopp won the UEFA Champions League after beating Tottenham Hotspur in the final and were long in the running to win their first ever English Premier League title. Moreover, several Liverpool players claimed a spot in the prestigious Ballon d'Or XI. Hence, we analyzed the mutual chemistry between the regular starters for Liverpool in their Premier League campaign.

Figure 1 shows the mutual offensive chemistry between the Liverpool players in the 2018/2019 English Premier League season, where green lines represent high mutual chemistry. We observe very strong offensive chemistry links among the three attackers as well as between the two full backs, Andrew Robertson and Trent Alexander-Arnold, and the attackers.

---

[8] https://github.com/SciSports-Labs/player-chemistry

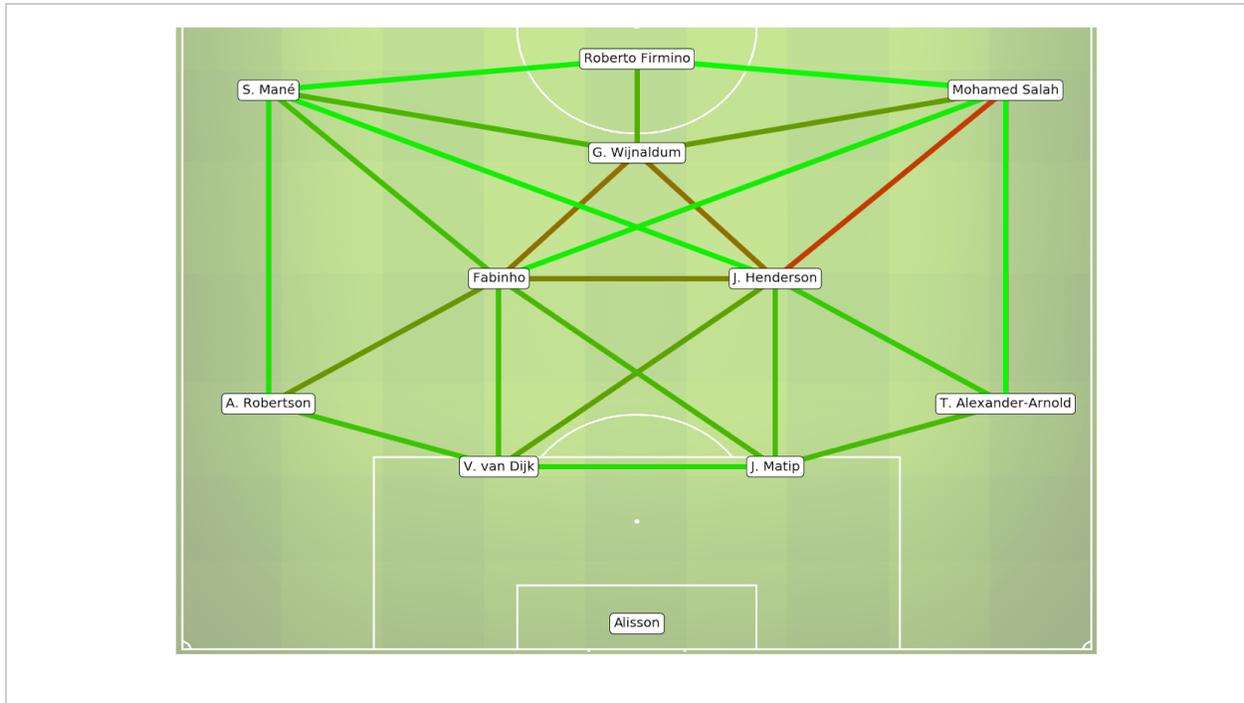

**Figure 1.** The mutual offensive chemistry between the Liverpool players in the 2018/2019 English Premier League season. Green lines reflect high offensive chemistry links, whereas dark red lines reflect low offensive chemistry links.

Figure 2 shows the mutual defensive chemistry between the Liverpool players in the 2018/2019 English Premier League season, where green lines represent high mutual chemistry. We observe strong defensive chemistry links in both midfield and central defense. Interestingly, the attackers also fulfill their defensive duties. In contrast, the defensive chemistry links between the central defenders and the full backs are rather weak. The links between Sadio Mané and Andrew Robertson on the left wing and Mohamed Salah and Trent Alexander-Arnold on the right wing have been omitted as these pairs of players have not been responsible for defending the same opponent for long enough. The link between Fabinho and Mohamed Salah has been omitted for the same reason.

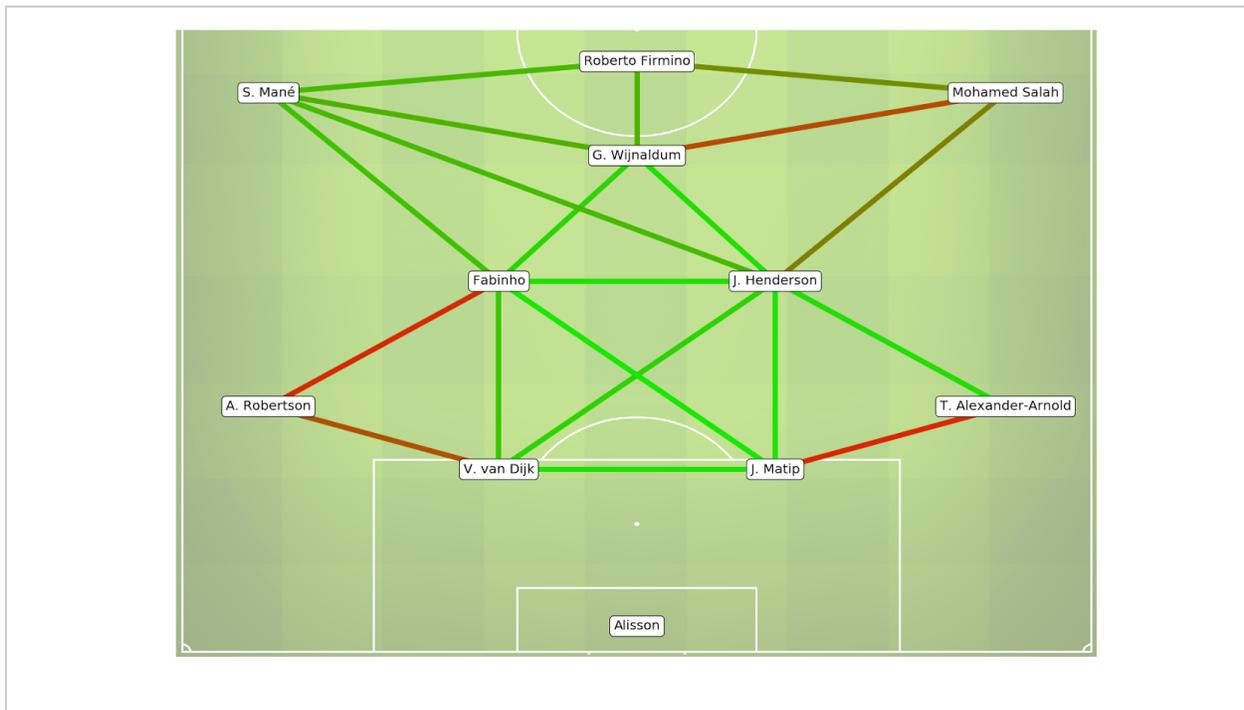

**Figure 2.** The mutual defensive chemistry between the Liverpool players in the 2018/2019 English Premier League season. Green lines reflect high defensive chemistry links, whereas dark red lines reflect low defensive chemistry links.

### 6.3. Mesut Özil's chemistry at Arsenal over time

Having joined Arsenal from Real Madrid at the start of the 2013/2014 season, Mesut Özil was a key player for *The Gunners* in the first few seasons after his arrival. However, over time, his performances have increasingly received criticism. Moreover, the German midfielder is even occasionally dropped from Arsenal's starting eleven. Hence, we analyze Özil's Joint Offensive Impact in the 2015/2016 through 2018/2019 seasons. As shown in Figure 3, the average JOI90 for the five best-ranked Arsenal player pairs has rapidly declined for both Özil and Arsenal since Alexis Sánchez' departure in 2018. Moreover, in the ongoing 2019/2020 Premier League season, the average JOI90 for the five best-ranked Arsenal player pairs cannot even match that of the average Premier League side. In summary, Arsenal's offensive chemistry has clearly declined over the years with Özil and his teammates making less of an offensive impact together.

Furthermore, Figure 4 shows the top-ten-ranked Arsenal player pairs in terms of Joint Offensive Impact per 90 minutes since the start of the 2015/2016 season. Mesut Özil features in six of these ten player pairs, including the top-four-ranked player pairs. Interestingly, the Özil - Sánchez partnership appears no less than three times in the top ten.

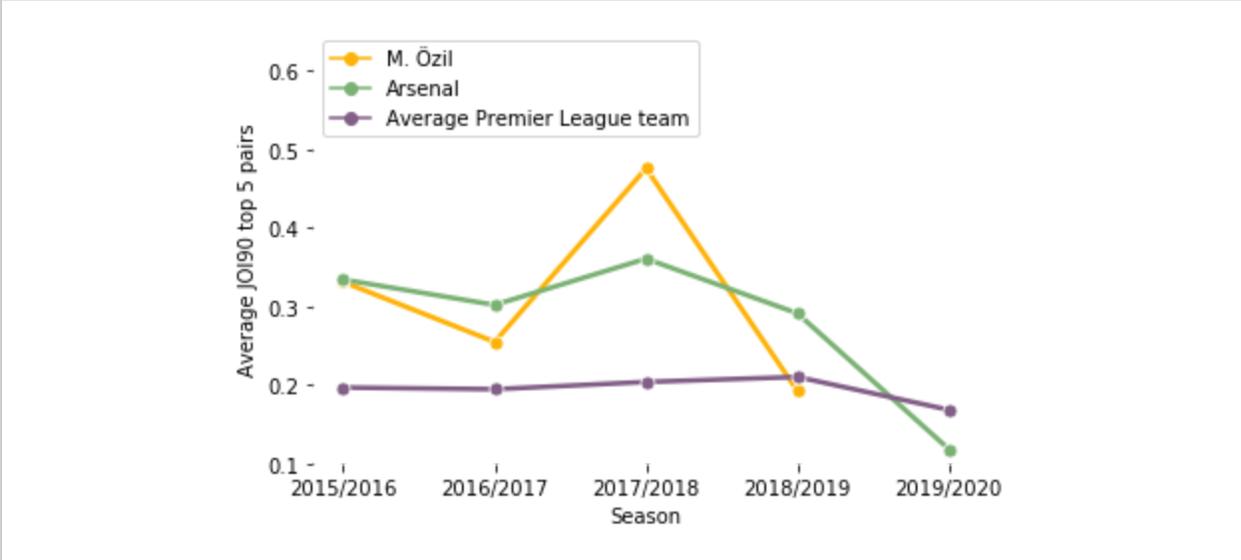

**Figure 3.** The evolution of the average Joint Offensive Impact per 90 minutes for Mesut Özil, Arsenal and the average Premier League team over the course of the past five seasons. Both Özil's and Arsenal's average have rapidly declined since Alexis Sánchez' departure in 2018.

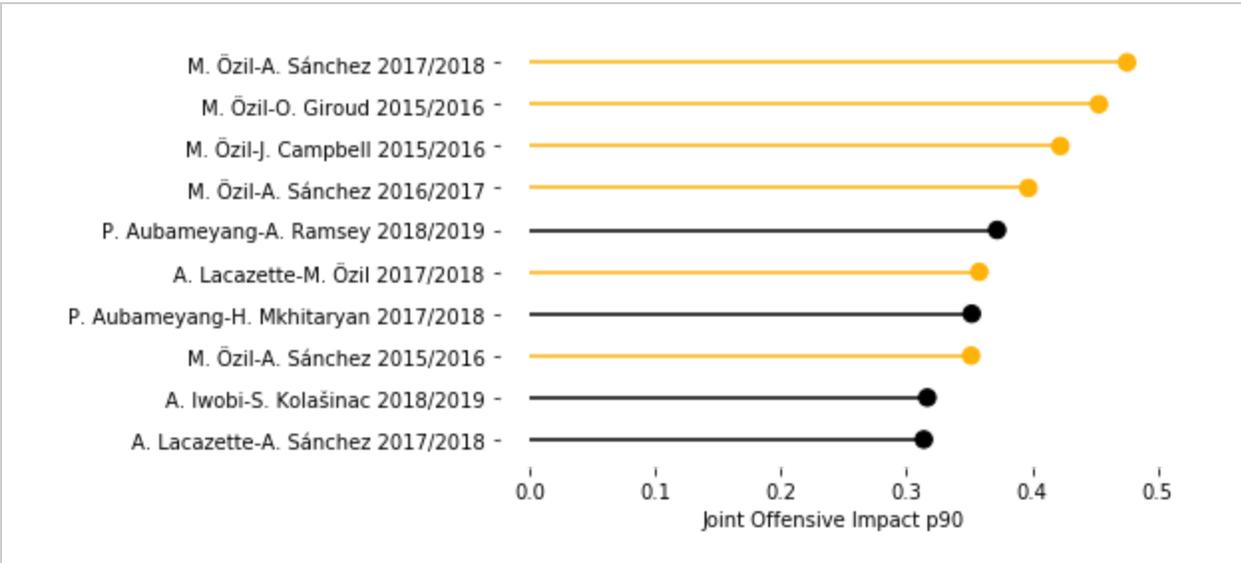

**Figure 4.** The top-ten-ranked Arsenal player pairs in terms of Joint Offensive Impact per 90 minutes since the start of the 2015/2016 season. Mesut Özil features in six of these ten player pairs, including the top-four-ranked pairs. Player pairs involving Mesut Özil are shown in orange.

# 7. Use cases

Our metrics for measuring and estimating offensive and defensive chemistry for pairs of players enable a multitude of use cases for practitioners within soccer clubs. In this section, we present three concrete use cases. First, we investigate which center back Manchester City should acquire taking chemistry with the players on their team into account. Second, we investigate which right winger Real Madrid should field from a chemistry perspective. Third, we investigate at which club Ajax' Hakim Ziyech would find optimal chemistry with the players surrounding him.

## 7.1. Which center back should Manchester City acquire?

Manchester City is left with only two senior options for the center back position in John Stones and Nicolás Otamendi after Vincent Kompany returned to his boyhood club Anderlecht in the summer of 2019 and Aymeric Laporte picked up a long-term knee injury just a few weeks after the summer transfer window in the English Premier League had closed. As a result, manager Pep Guardiola has been forced to play Fernandinho, who was pivotal to City's midfield in their last few campaigns, at center back. Unsurprisingly, several high-profile center backs have been linked with a move to the Etihad Stadium during the first half of the 2019/2020 season, including Tottenham Hotspur's Toby Alderweireld, Napoli's Kalidou Koulibaly, and Internazionale's Milan Škriniar. Hence, we investigate which of these potential targets would be the best fit in terms of chemistry.

To address this use case, we predict the defensive and offensive chemistry between each of the three potential targets and a selection of relevant Manchester City players. In terms of defensive chemistry, Koulibaly would gel best with center back Stones as well as defensive midfielders Fernandinho and Rodrigo, whereas Alderweireld would gel best with Otamendi, as can be seen in Figure 5. In terms of offensive chemistry, Koulibaly would fit best with all four players under consideration, as can be seen in Figure 6. However, with respect to offensive players Kevin De Bruyne, David Silva, Bernardo Silva, Riyad Mahrez, Gabriel Jesus and Sergio Agüero, Alderweireld would be a better choice due to his strong long-passing skills, as can be seen in Figure 7. The high offensive chemistry between Alderweireld and De Bruyne is partly due to the 41 matches that they have played together for the Belgian national team. Hence, Koulibaly would be the most obvious pick from a defensive perspective, whereas Alderweireld would be the most obvious pick from an offensive perspective. Although Škriniar would be the least obvious choice overall, he would be a better fit with City's offensive compartment in terms of offensive chemistry than Koulibaly.

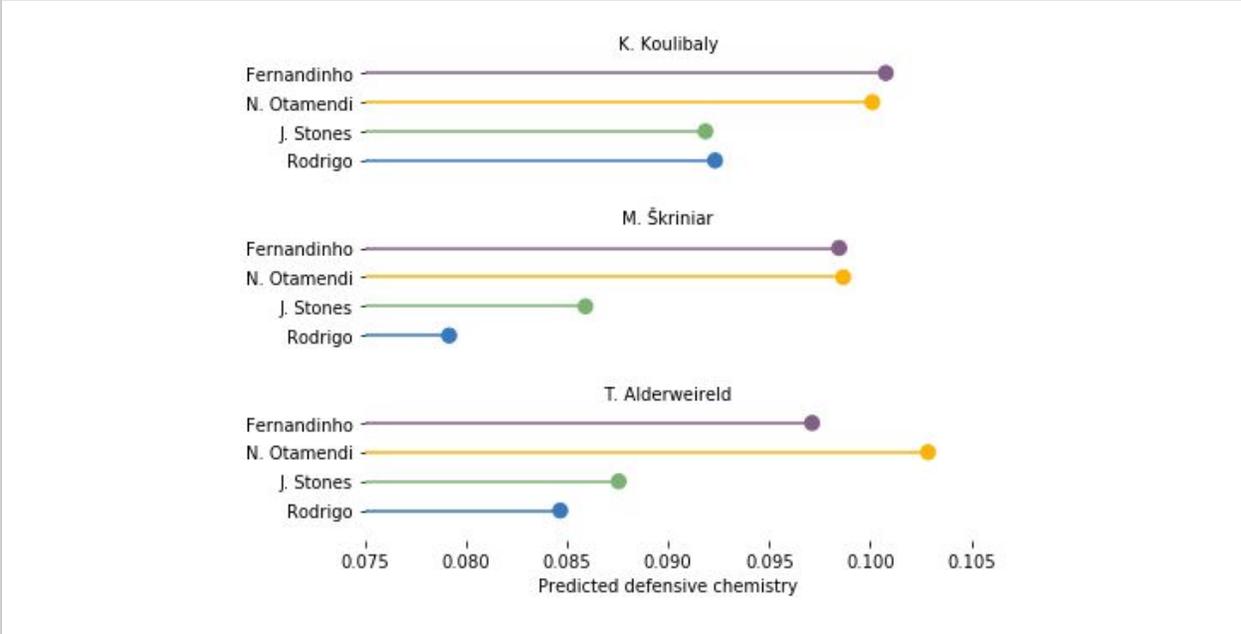

**Figure 5.** Predicted defensive chemistry between potential targets Koulibaly, Škriniar and Alderweireld on one hand and Fernandinho, Otamendi, Stones and Rodrigo on the other hand. In terms of defensive chemistry with the defensive compartment, Koulibaly would be the best fit with all players but Otamendi.

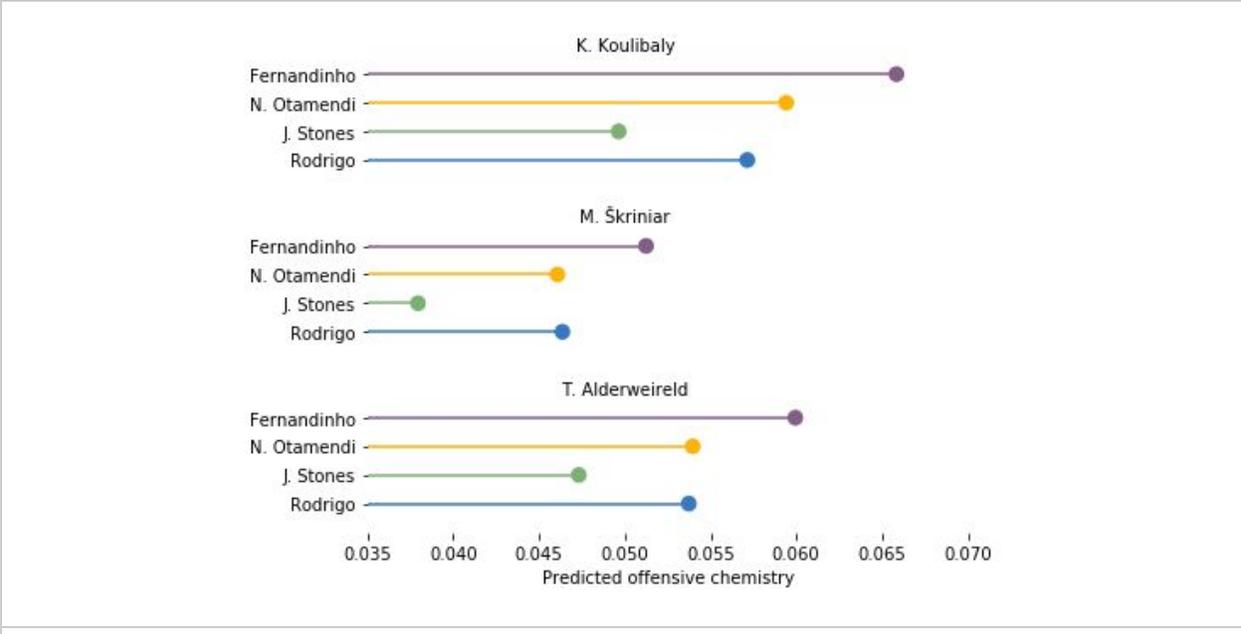

**Figure 6.** Predicted offensive chemistry between potential targets Koulibaly, Škriniar and Alderweireld on one hand, and Fernandinho, Otamendi, Stones and Rodrigo on the other hand. In terms of offensive chemistry with the defensive compartment, Koulibaly would be the best fit.

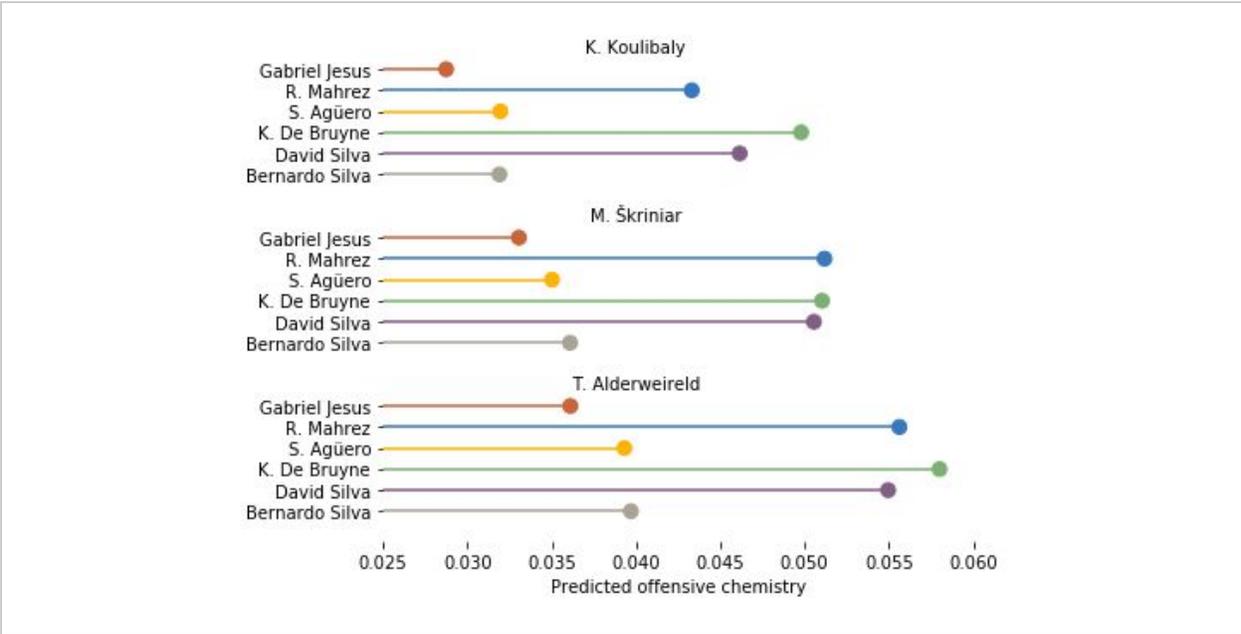

**Figure 7.** Predicted offensive chemistry between potential targets Koulibaly, Škriniar and Alderweireld on one hand, and Jesus, Mahrez, Agüero, De Bruyne, D. Silva and B. Silva on the other hand. In terms of offensive chemistry with the offensive compartment, Alderweireld would be the best fit.

## 7.2. Which right winger should Real Madrid field?

Despite signing elite players such as Eden Hazard, Luka Jovic, Rodrygo, Vinícius Júnior and Mariano Diaz for a combined transfer fee of over 300 million dollar, Real Madrid boss Zinedine Zidane is still struggling to rebuild his side following the departure of Portuguese star player Cristiano Ronaldo to Juventus in the summer of 2018. The French manager has been particularly struggling with filling the spot on the right wing as 30-year-old Gareth Bale's form has been on the decline and Lucas Vázquez picked up an early-season toe injury that has kept him sidelined for several months. Hence, we investigate whether Gareth Bale, Rodrygo or Vinícius Júnior would be the best fit for the spot on the right wing from a chemistry perspective, both in the short run and in the long run.

To address this use case, we compute the average predicted offensive chemistry between Bale, Rodrygo and Vinícius Júnior on one hand, and all other Real Madrid players on the other hand. As shown in Figure 8, Bale had the highest average predicted offensive chemistry at the start of the 2019/2020 season with Rodrygo ranking second and Vinícius Júnior ranking third. However, when we artificially increase the number of matches played with each of the other Real Madrid players over time, the story changes. That is, we expect the average predicted offensive chemistry for Rodrygo and Vinícius Júnior to increase rapidly, whereas the average predicted offensive chemistry for Bale will increase much slower. Vinícius Júnior is expected to overtake Bale around the 20-extra-matches mark, whereas Rodrygo is expected to do the same around the 40-extra-matches mark.

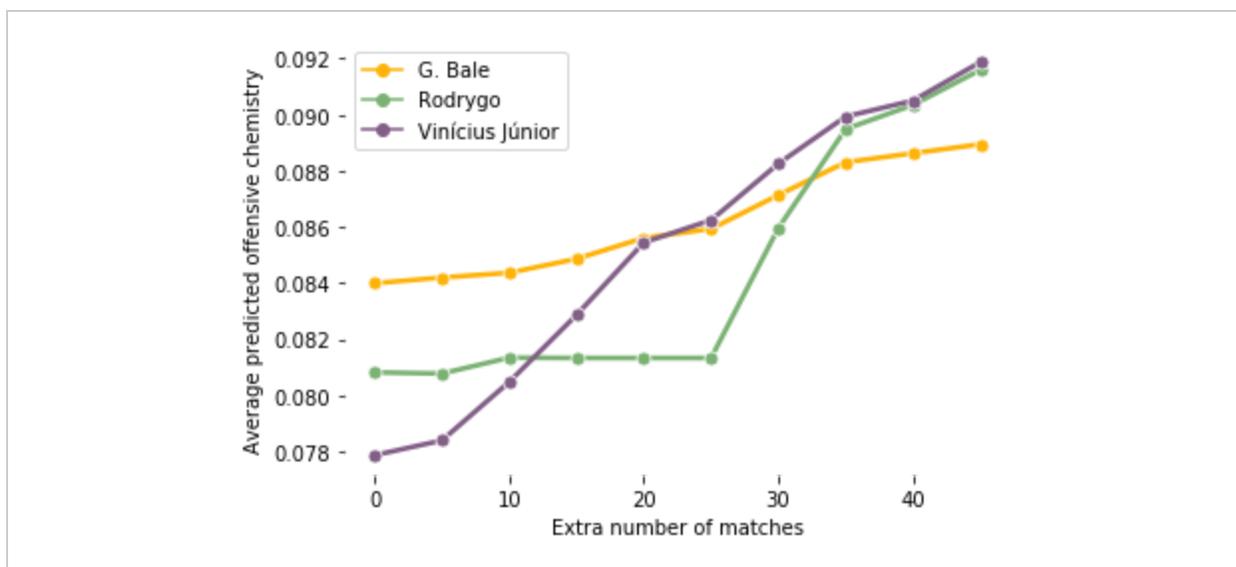

**Figure 8.** The average predicted offensive chemistry between Bale, Rodrygo and Vinícius Júnior on one hand, and all other Real Madrid players on the other hand. While Bale had the highest offensive chemistry at the start of the 2019/2020 season, he is expected to be quickly overtaken by both Rodrygo and Vinícius Júnior as they collect more matches.

However, when comparing the mutual offensive chemistry between each of Bale, Rodrygo and Vinícius Júnior on one hand, and Benzema, Kroos, Hazard and Isco on the other hand, Bale turns out the have the highest mutual offensive chemistry with all players but latest signing Hazard, as is shown in Figure 9. In contrast, the Belgian winger is expected to gel better with both Rodrygo and Vinícius Júnior.

In summary, based on our metrics, we would advise Real Madrid boss Zinedine Zidane to give both Rodrygo and Vinícius Júnior enough playing time in the 2019/2020 season to improve their chemistry with their teammates. Furthermore, as Eden Hazard is set to become Real Madrid's next star after the departure of Cristiano Ronaldo to Juventus, it is essential for him to gel well with his teammates. The Belgian winger seems to gel better with Rodrygo than with Bale.

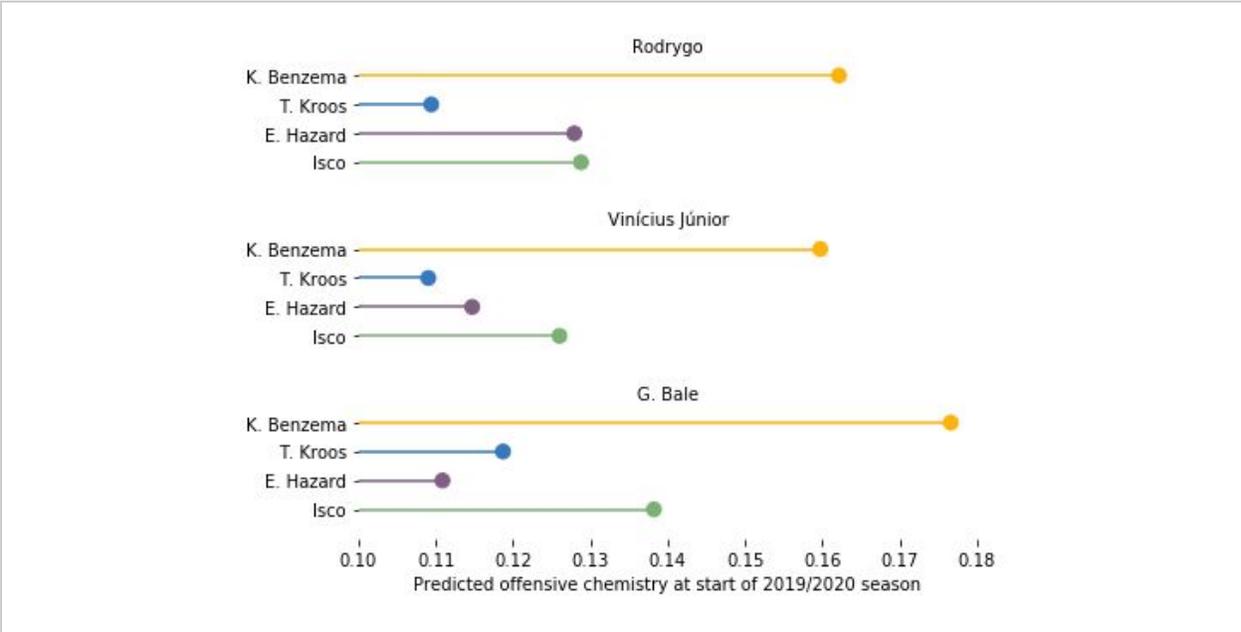

**Figure 9.** The mutual offensive chemistry between each of Bale, Rodrygo and Vinícius Júnior on one hand, and Benzema, Kroos, Hazard and Isco on the other hand. Bale has the highest mutual offensive chemistry with all players but Hazard.

### 7.3. Which club should Hakim Ziyech move to?

Outclassing the opposition in the Dutch Eredivisie week after week, attacker Hakim Ziyech is set to move to an elite European league in the summer of 2020. After an incredible 2018/2019 season, which saw Ajax reach the Champions League semi-finals, several clubs showed interest in signing the Moroccan international in the summer of 2019. With a multitude of options on the table for the talented attacker, we step in Ziyech's shoes and help him evaluate his options. In particular, we help the attacker find a team where he would experience optimal chemistry with the players surrounding him, while at the same time being able to collect sufficient playing time as well as being able to further develop himself.

To address this use case, we perform two steps. In the first step, we identify the five most appropriate clubs for Hakim Ziyech in terms of playing time, future development and likeliness that an agreement can be reached between club and player. To do so, we first use SciSports' machine-learned models that predict the number of minutes Ziyech would play for each candidate club, by how many points his SciSkill would increase or decrease in the next year at each candidate club, and the likeliness that each candidate club can afford to acquire Ziyech from Ajax and pay his wages. Next, we transform each of the three predictions into a score on a zero-to-one scale and compute the average score for each candidate club. Finally, we sort the candidate clubs according to their average score in descending order and select the top-five-ranked clubs: Internazionale, Roma, Chelsea, Bayern Munich, and Arsenal.

In the second step, we first predict the offensive chemistry for all possible player pairs within the selected teams as well as between Hakim Ziyech and each of the players on the selected teams, and then use our Team Builder to assemble the maximum-chemistry line-up for each of the five selected clubs. We force the Team Builder to include Hakim Ziyech and to select precisely ten players from the squad of each selected club by imposing the following additional constraint:

$$\sum_{p \in P_T} x_p = 10,$$

where $P_T$ is the set of all players in the current squad of candidate club T. Furthermore, we set the weighting parameter $\alpha$ to 1 to only consider the offensive chemistry.

Given that we are searching for the best-suited club from Ziyech's perspective, we investigate in which of the five optimal-chemistry teams that our Team Builder assembled, Ziyech would experience the highest average offensive chemistry with his teammates. As is shown in Figure 10, the Moroccan attacker would achieve the highest offensive chemistry with the players who would be surrounding him at Bayern Munich. Figure 11 shows how Bayern Munich's optimal-chemistry side would look like when they would have Ziyech in their ranks.

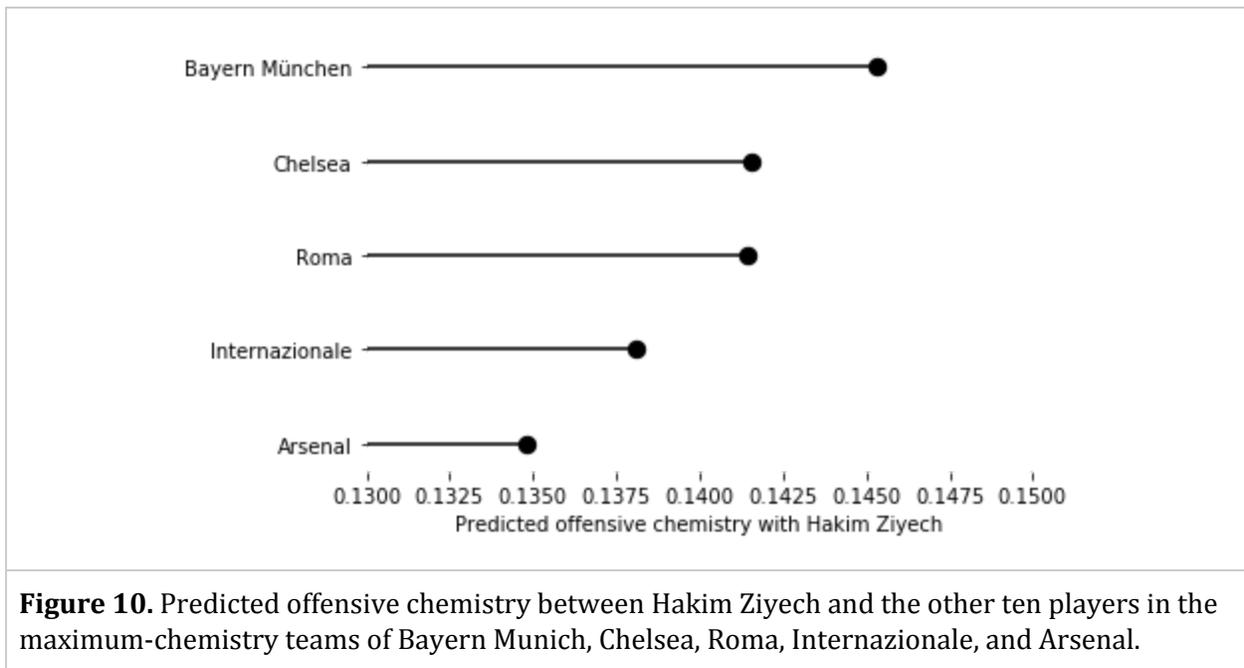

**Figure 10.** Predicted offensive chemistry between Hakim Ziyech and the other ten players in the maximum-chemistry teams of Bayern Munich, Chelsea, Roma, Internazionale, and Arsenal.

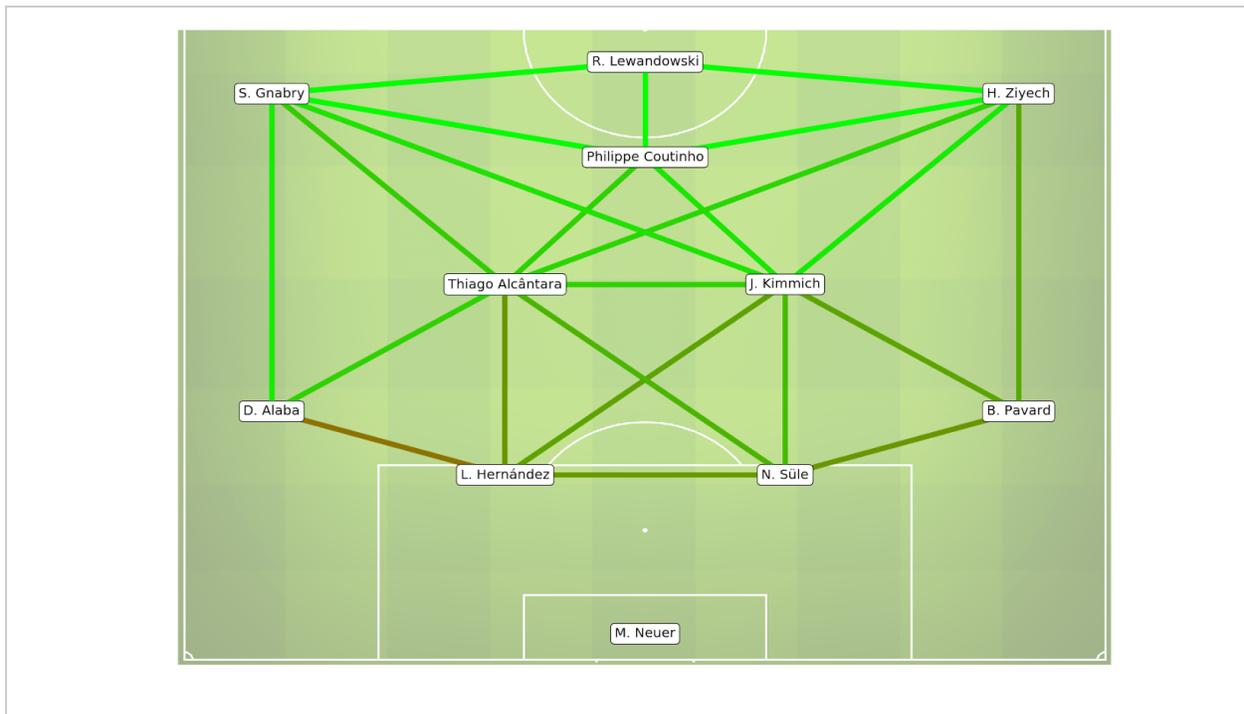

**Figure 11.** The mutual offensive chemistry between the Bayern Munich players for the maximum-chemistry team that includes Hakim Ziyech.

## 8. Conclusions

This paper has taken a first step towards providing insight into how well a team of soccer players gels together. We presented metrics that capture the mutual offensive and defensive chemistry for a pair of players by quantifying their joint impact on scoring goals and preventing their opponents from scoring goals. We introduced two machine-learned models that predict these metrics for players who have never played together. Furthermore, we introduced a Team Builder that assembles a maximum-chemistry team from a given set of players. We demonstrated how our chemistry metrics and Team Builder can be used to identify an appropriate transfer target for a club, to decide on the best possible line-up for a team, and to provide transfer advice to a player.

In the future, we plan to pursue several different avenues to further investigate the chemistry among the players on a soccer team. We aim to expand our metrics and models from capturing the mutual chemistry between pairs of players to capturing the mutual chemistry among groups consisting of more than two players. Furthermore, we aim to incorporate spatio-temporal tracking data to obtain more accurate estimates for our defensive chemistry metric.

## Acknowledgements

We thank our SciSports colleagues Max Adema and Mick Bosma as well as Simões Anderson, Joris Bekkers, Pieter Robberechts and Robert Seidl for their comments and suggestions that helped improve this paper. We also thank Wyscout for supplying the match event data used in this paper.

## References

[1] R. Beal and S. Ramchurn, "Learning the Value of Teamwork to Form Efficient Teams," in *Proceedings of the 2019 StatsBomb Innovation in Football Conference*, London, United Kingdom, 2019.

[2] P. Robberechts, J. Van Haaren, L. Bransen, and J. Davis, "Contextualized Performance Projections for Soccer Players," presented at the 2020 OptaPro Analytics Forum, London, United Kingdom.

[3] M. Goldberg, "Evaluating Lineups and Complementary Play Styles in the NBA," Harvard College, Cambridge, Massachusetts, United States, 2017.

[4] J. Kuehn, "Accounting for Complementary Skill Sets When Evaluating NBA Players' Values to a Specific Team," in *Proceedings of the 2016 MIT Sloan Sports Analytics Conference*, Boston, Massachusetts, United States, 2016.

[5] A. Maymin, P. Maymin, and E. Shen, "NBA Chemistry: Positive and Negative Synergies in Basketball," *Int. J. Comput. Sci. Sport*, 2013.

[6] T. Decroos, L. Bransen, J. Van Haaren, and J. Davis, "Actions Speak Louder Than Goals: Valuing Player Actions in Soccer," in *Proceedings of the 25th ACM SIGKDD International Conference on Knowledge Discovery and Data Mining*, Anchorage, Alaska, United States, 2019.

[7] B. Aalbers, "SciSkill Index - Why and How," *SciSports*, 2016. [Online]. Available: https://www.scisports.com/sciskill-index-why-and-how/.

[8] B. Aalbers and J. Van Haaren, "Distinguishing Between Roles of Football Players in Play-by-play Match Event Data," in *Proceedings of the 5th Workshop on Machine Learning and Data Mining for Sports Analytics*, Dublin, Ireland, 2018.

[9] D. Cervone, A. D'Amour, L. Bornn, and K. Goldsberry, "POINTWISE: Predicting Points and Valuing Decisions in Real Time with NBA Optical Tracking Data," in *Proceedings of the 2014 MIT Sloan Sports Analytics Conference*, Boston, Massachusetts, United States, 2014.

[10] L. Prokhorenkova, G. Gusev, A. Vorobev, A. V. Dorogush, and A. Gulin, "CatBoost: Unbiased Boosting with Categorical Features," in *Advances in Neural Information Processing Systems*, 2018, pp. 6638–6648.